\documentclass[prl,preprint,aps,superscriptaddress]{revtex4}
\usepackage{amsmath}
\usepackage{graphicx}
\usepackage{tabularx}
\usepackage{amssymb}
\usepackage{hyperref}
\usepackage[usenames]{color}
\definecolor{rltred}{rgb}{0.75,0,0}
\definecolor{rltgreen}{rgb}{0,0.5,0}
\hypersetup{colorlinks,linkcolor=rltred,citecolor=rltgreen}

\begin{document}

\title{Engineering the initial state in broadband population inversion}

\author{Bo Y. Chang}
\affiliation{School of Chemistry (BK21$^+$), Seoul National University, Seoul 151-747, Republic of Korea}

\author{Seokmin Shin}
\affiliation{School of Chemistry (BK21$^+$), Seoul National University, Seoul 151-747, Republic of Korea}

\author{Ignacio R. Sola}
\affiliation{Departamento de Qu\'imica F\'isica, Universidad Complutense, 28040 Madrid, Spain}
\email{isola@quim.ucm.es}

\begin{abstract}
Quantum systems with sublevel structures prevent full population inversion 
from one manifold of sublevels to the other using strong ultrafast resonant 
pulses.  In this work we explain the mechanism by which this
population transfer is blocked. We then develop a novel concept of geometric
control, assuming full or partial coherent manipulation within the manifolds
and show that by preparing specific coherent superpositions in the initial
manifold, full population inversion or full population blockade, {\it i.e}
laser-induced transparency, can be achieved.
In particular, by parallel population transfer we show how population inversion
between the manifolds can be obtained with minimal pulse area. As the number
of sublevels increases, population inversion can overcome the pulse area 
theorem at the expense of full control over the initial manifold of sublevels.

\end{abstract}


\maketitle

In this work we are concerned with intrinsic properties of the dynamics
of systems with manifolds of sublevels, described by two (or more) quantum
numbers, that hereafter will be generically called quantum structures.
From the point of view of controlling the system dynamics, quantum
structures pose several interesting problems.
Quantum control\cite{QC} typically implies the ability to manipulate 
interfering pathways, which increases with the number of levels 
that participate in the dynamics as long as the system is {\em controllable} 
\cite{controllability}. 
A multi-level structure would therefore offer more control opportunities
at the expense of the ability to manipulate within the substructures.
Our general goal is to investigate whether quantum structures
limit, or conversely help, in controlling the system. In this paper we will be 
concerned with coarse-grained goals, where the objective of the control 
will be the state of the manifold given by the first quantum number,
not the detailed state of the sublevels. 
In finding the best possible controls we will {\em assume} that the
substructure is partially controllable, that is, that given some 
constraints, any possible wave function within a subset of the sublevels 
can be prepared \cite{wfcontrol}.
Building on this assumption we will develop a {\em geometric control} approach
that allows finding the optimal initial wave functions that maximize
the yield of the desired process. This procedure does not prescribe an
optimal field, but implicitly assumes that a field can be found, and
makes full use of the quantum structures.

Let us consider a simple and very general process in systems with
a congested spectrum: absorption
from the ground, initial manifold to the excited, target manifold,
by means of a strong ultrashort pulse, with a bandwidth much larger than
the energy spacing of the sublevels within the fine structure.
As a molecular example, we can conceive controlling an 
electronic transition using a broadband pulse, where the goal is to invert the 
population to the 
excited state, regardless of the vibrational populations\cite{molpipulse}.
Although one may think that the substructure, specially in the case of very 
different associated time-scales, does not affect the 
overall transition, in the strong field case the opposite occurs.
The unpopulated levels of the fine structure induce Stark shifts\cite{Starkshift}
and create effective detunings from the resonance that limit the
extent of the Rabi oscillations. That is, assuming
that all the different sublevels are dipole allowed, regardless
of the strength of the pulse, the maximum population that can be
reached is typically much smaller than one.

A simple theoretical model explains this observation.
Let us first assume that the different sublevels within each manifold are
degenerate, $\Delta E \ll \Delta \omega \approx 0$, and that all transient
dipoles are equal.
Then the equations of motion for every sublevel of the excited manifold 
$|e,k\rangle$, and for every sublevel in the initial manifold 
$|g,j\rangle$, are the same. For a resonant transition:
$\dot{a}_j = i \Omega(t) \sum_k b_k(t)/2$ and
$\dot{b}_n = i \Omega(t) \sum_j a_j(t)/2$,
where $\Omega(t)$ is the Rabi frequency.
However, the initial conditions in the manifold $|g,j\rangle$, where
we assume a single state, $|g,1\rangle$, is initially populated, 
break the symmetry so that three different probability amplitudes
describe the dynamics. We write,
\begin{equation}
|\Psi(t)\rangle = a(t) | g, 1\rangle + \bar{b}(t) \sum_k^{N_f} |e, k\rangle +
\bar{c}(t) \sum_{j'}^{N_i-1} |g, j' \rangle
\end{equation}
where the prime indicates that the initial state is excluded from the 
summation in the $|i,j\rangle$ manifold, 
and $\bar{b}(t)$ and $\bar{c}(t)$ are mean probability amplitudes,
which in fact behave {\em exactly} as every sublevel amplitude.
We define now the collective excited $|E\rangle$
and Raman $|R\rangle$ states
\begin{equation}
|E\rangle = \frac{1}{\sqrt{N_f}} \sum_k^{N_f} |e, k\rangle
\end{equation}
\begin{equation}
|R\rangle = \frac{1}{\sqrt{N_i-1}} \sum_{j'}^{N_i-1} |g, j'\rangle
\end{equation}
which together with the initial state $|i\rangle \equiv |g,1\rangle$ form
an orthonormal basis such that
$|\Psi(t) \rangle = a(t) | i\rangle + B(t) |E\rangle + C(t) |R\rangle$
with $B(t) = \sqrt{N_f}\, \bar{b}(t)$, $C(t) = \sqrt{N_i-1} \bar{c}(t)$,
and $|a(t)|^2+|B(t)|^2+|C(t)|^2 = 1$. 
The effective Hamiltonian in this basis is
\begin{equation}
{\sf H} := -\frac{1}{2}\sqrt{N_f}\Omega(t) \left( |i\rangle\langle E| +
\sqrt{N_i -1} |E\rangle\langle R| + \mathrm{c.c.} \right)
\end{equation}
with simple analytic eigenvalues and eigenvectors ({\it aka} dressed states)
\cite{Shore}. 
When $N_i = 2$ the wave function dynamics is particularly interesting,
with a population in the manifold of excited states given by
\begin{equation}
P_E(t) = \sum_k^{N_f} | \langle e,k | \Psi(t)\rangle |^2 = 
|\langle E | \Psi(t) \rangle |^2 = \frac{1}{2} \sin^2 \left(
\sqrt{\frac{N_f}{2}}\theta(t) \right)  \label{PE2}
\end{equation}
where $\theta(t) = \int_{-\infty}^{t} \Omega(t')dt'$, such that
$\theta(\infty)$ is the pulse area, ${\cal A}$. 
A maximum of $50$\% population can reach the excited state, whereas there
is Rabi flopping (at twice the period of oscillation) between the initial 
state $|i\rangle$ ($|i,1\rangle$) and $|R\rangle$ ($|i,2\rangle$), 
\begin{equation}
P_R(t) = | \langle R | \Psi(t)\rangle|^2 = \sin^4 \left( \frac{1}{2}
\sqrt{\frac{N_f}{2}} \theta(t) \right)   \label{PR}
\end{equation}
{\it i.e.} there is a very efficient Raman Stokes transition.
Increasing the number of sublevels in the initial manifold only blocks the 
population transfer more efficiently. 
For large $N_i$, 
we obtain a population in the excited manifold of
\begin{equation}
P_E(t) = \frac{1}{N_i} \sin^2 \left( \frac{1}{2}
\sqrt{N_fN_i} \theta(t) \right) \label{PE}  
\end{equation}
Unlike in $N_i = 2$, for large $N_i$ the maximum 
Raman populations depend also on $N_i^{-1}$. 
Larger pulse areas only increase the frequency of the oscillations, but the 
initial state is mostly decoupled. 
The population dynamics behave as in an off-resonant excitation,
with an effective detuning created by the Autler-Townes splitting\cite{AutlerTownes} between 
the $|E\rangle$ and $|R\rangle$ states, coupled by an $\sqrt{N_i - 1}$
stronger Rabi frequency than the initial state with the $|E\rangle$ state.

\begin{figure}
\includegraphics[width=6.5cm,scale=1.0]{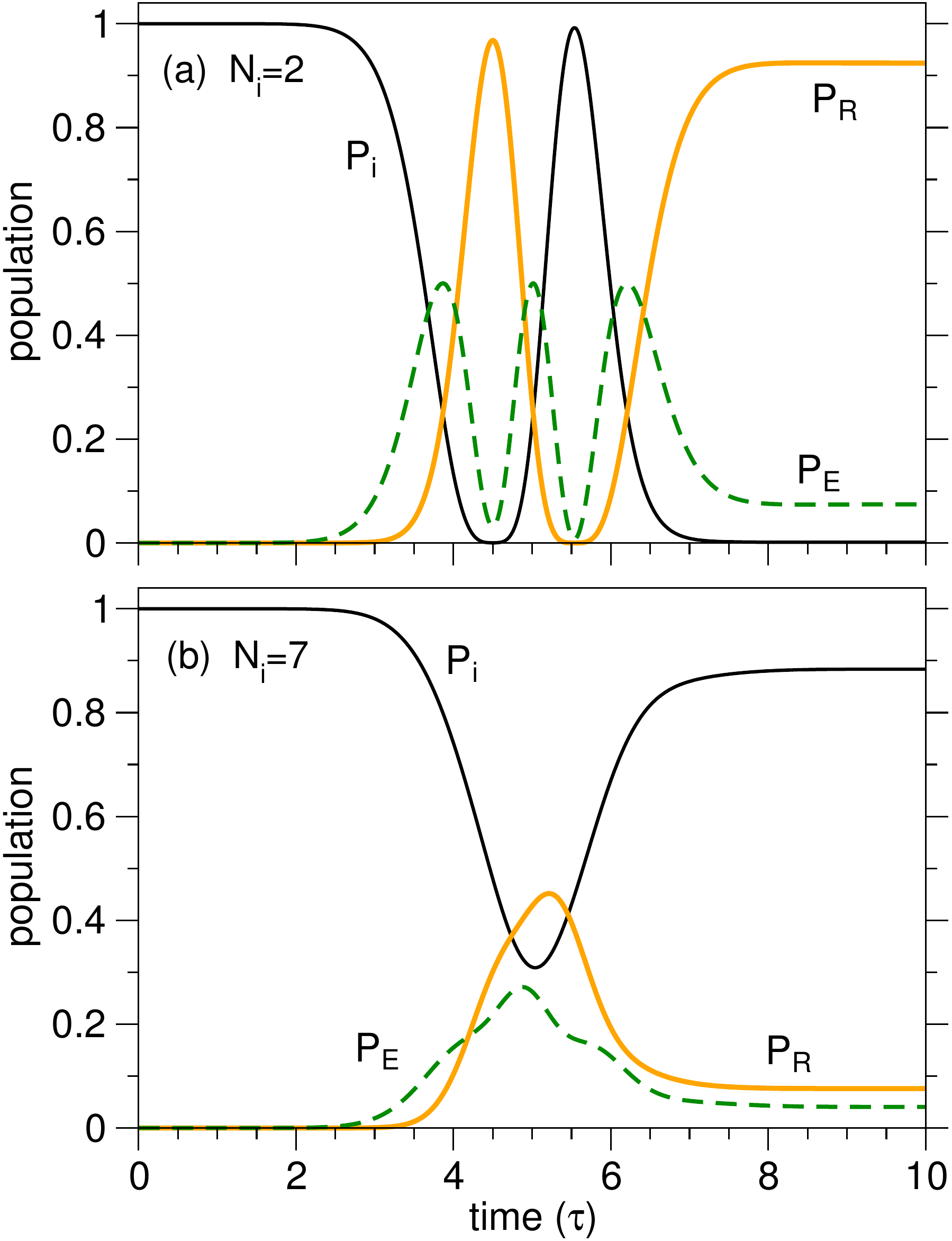}
\caption{Population histories for a system with (a) $N_i = 2$ ($N_f=1$), (b) 
$N_i = 7$ ($N_f=7$). In both cases the pulse area is ${\cal A} = 5\pi$
and the energy difference between the levels is $\Delta E = 0.4 \tau^{-1}$.}
\end{figure}

Fig.1 shows the population dynamics for different sublevel structures,
using time-scaled units. Case 1 refers to $N_i = 2$, $\Delta E = 0.4\tau^{-1} < 
\Delta \omega = 4\ln{2}$ (where $\Delta \omega$ is the bandwidth of the 
Gaussian pulse, with scaled width, $\tau = 1$), 
$N_f = 1$ and the peak Rabi frequency is $\Omega_0 = 2$.
In case 2 we use $N_i = N_f = 7$ with the same energy splitting and pulse
parameters as before. The results show (i) efficient Raman transfer
for the first case and (ii) population locking in the second case, 
qualitatively in agreement with Eqs.(6) and (7) even for non-degenerate 
structures. The main effect provoked by the energy splittings is to
allow more population flow to the most excited sublevels of the initial and
final manifolds, because the energy difference partially off-sets the 
effective detuning. But the effect is too small to qualitatively change the
dynamics.

Is it possible to optimize the pulse parameters to increase
the efficiency of the population transfer? 
Clearly, as long as the initial state is a single sublevel Eq.(\ref{PE}) 
limits the maximum population that can be transferred using transformed-limited
pulses. 
The control requires manipulation of the initial wave function.
We will assume that the initial manifold can be manipulated before
$\Omega(t)$ acts such that we have full controllability within a 
given subset of states. 
Typically this requires the use of laser pulses of very different frequencies.
In molecular physics, several control schemes have been proposed that
imply creating coherences in the initial electronic state by means
of infrared pulses before the optical field is used\cite{IRcontrol}.
However, instead of explicitly finding these pulses, in this work 
we will develop a geometrical approach.

We want to maximize the population on the final manifold
at time $T$, given by the functional ${\cal F}$, 
\begin{equation}
{\cal F} = \sum_m^{N_f} \langle \Psi(t_i)| {\sf U}(t_i,T;\Omega) | e,m \rangle
\langle e, m | {\sf U}(T,t_i;\Omega) | \Psi(t_i) \rangle \label{Fop}
\end{equation}
for fixed $\Omega(t)$, with respect to changes in the initial wave function 
$|\Psi(t_i)\rangle$, where $t_i$ is the initial time.
This amounts to finding 
${\sf U}(t_i,0;\Omega_i)|g,1\rangle = |\Psi(t_i)\rangle$. 
Instead of explicitly finding the new field, we assume controllability
and use a variational approach to simply obtain the rotation matrix 
${\sf R}_i |g,1\rangle = |\psi_i\rangle = \sum_j^{N_c} a_{ij} |g,j\rangle$
(where the sum can be constrained to a subset of the levels of the
initial manifold, {\it i.e.} $N_c \le N_i$) such that ${\cal F}$ is maximal. 
We therefore substitute
$|\Psi(t_i)\rangle$ by $|\psi_i\rangle$ in Eq.(\ref{Fop}).
The optimization
is purely geometrical and one can use the Rayleigh-Ritz approach.
Thus we construct the matrix ${\sf F}$ with elements
\begin{equation}
{\sf F}_{jk} = \sum_m \langle i, j | {\sf U}(0,T;\Omega) | f, m \rangle 
\langle f, m | {\sf U}(T,0;\Omega) | i, k \rangle  \label{Fij}
\end{equation}
Restricting $|\psi_i\rangle$ to be normalized 
(equivalently, ${\sf R}_i$ to be unitary) we obtain the secular equation, 
${\sf F} |\psi_i\rangle = \chi_i |\psi_i\rangle$.
The solutions are the eigenvectors of ${\sf F}$ which give the 
yields of population transfer $\chi_i$.

\begin{figure}
\includegraphics[width=8.0cm,scale=1.0]{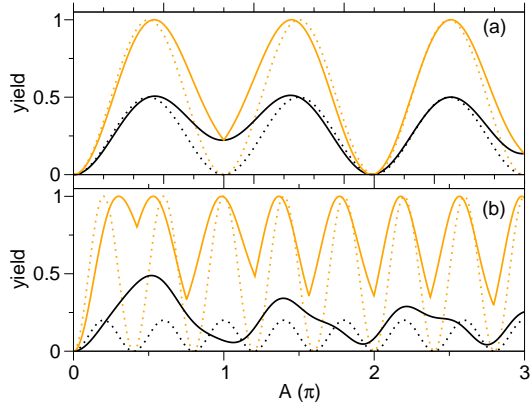}
\caption{Maximum population transfer to the excited manifold at final time
as a function of pulse area for a system with (a) $N_i = 2$ and (b) $N_i = 5$.  
In both cases $N_f = N_i$. Dotted lines represent results for degenerate
structures while solid lines give results for structures with energy
spacings $\Delta E = 0.4 \tau^{-1}$.}
\end{figure}

In Fig.2 we compare the results of the optimization with the yields 
obtained from the initial state $|i\rangle$ for degenerate structures
$\Delta E = 0$ and with constant energy spacing between adjacent
levels $\Delta E = 0.4 \tau^{-1}$, for two different manifolds,
$N_c = N_i = N_f = 2$ and $N_c = N_i = N_f = 5$.
The optimal yields for degenerate structures follow the pattern
of Rabi oscillations coinciding with the population transfer from the initial
state, but with full population transfer at multiples of $\pi$ of the pulse
area. For non-degenerate structures the minima of these oscillations does
not drop to zero but {\em increases} with the number of levels, despite
the $N^{-1}$ factor in Eq.(\ref{PE}). 

It is interesting to analyze generic features of the optimized initial states.
When the extended pulse area ${\cal A}_e = \sqrt{N_i N_f} 
{\cal A}$ is an odd multiple of $\pi$, the optimal initial states have a 
very clear structure: All their coefficients are equal.
This result can be explained analytically for the degenerate structure,
following exactly the same steps as in Eq.(1). We now define
\begin{equation}
|\Psi(t)\rangle = \bar{a}(t) \sum_j^{N_p} | g, 1\rangle + \bar{b}(t) 
\sum_k^{N_f} |e, k\rangle + \bar{c}(t) \sum_{j}^{N_u} |g, j \rangle
\end{equation}
where $\bar{a}(t)$ is the mean amplitude of all $N_p$ initially
populated levels in the initial manifold, whereas $N_u = N_i - N_p$
is the set of unoccupied states. 
Together with the previously defined $|R\rangle$ and $|E\rangle$ collective
states, the collective initial state
\begin{equation}
|I\rangle = \frac{1}{\sqrt{N_p}} \sum_{j}^{N_p} |g, j\rangle
\end{equation}
forms a orthonormal set such that the Hamiltonian can be written as
\begin{equation}
{\sf H} := -\frac{1}{2}\sqrt{N_f}\Omega(t) \left( \sqrt{N_p} |I\rangle
\langle E| + \sqrt{N_u} |E\rangle\langle R| + \mathrm{c.c.} \right)
\end{equation}
with the same eigenvalues as before. 
Given the wave function $|\Psi(t)\rangle = A(t) |I\rangle + B(t) |E\rangle
+ C(t) |R\rangle$ with $A(t) = \sqrt{N_p}\, \bar{a}(t)$,
the probability of reaching the excited manifold is
\begin{equation}
P_E(t) = \frac{N_p N_i}{N_p^2+N_u^2} \sin^2\left( \frac{1}{2} {\cal A}_e \right)
\end{equation}
Whenever ${\cal A}_e(t)$ is an odd multiple of $\pi$, full population inversion 
can be achieved if {\em all} the sublevels of the initial manifold are 
equally populated and {\em in phase}. For different energy spacings other choices
of phases and populations give better results, but the populations are
always almost equal. On the other hand,
it is simple to proof that when the initial probability amplitudes are all 
{\em out of phase} 
             (such that $\sum_j^{N_p} a_j(0) = 0$) then ${\cal F}$ is
minimized and perfect transparency can be achieved, {\em i.e.} the 
population in the excited manifold is zero at all times\cite{LIT}. 
Similar results are obtained with nondegenerate structures although
the transparency is no longer perfect.
Since there are many more possible solutions that minimize the yield
(exactly $N_i -1$ orthogonal eigenvectors for the degenerate structure) 
than those that maximize
the yield (a single solution for the degenerate case) the set of eigenvalues
fills from below and the subspace of population transfer is of very small
dimension. From the point of view
of quantum controllability, population transfer is a difficult problem.
Nevertheless, other yields greather than zero can be achieved 
when $\Delta E \neq 0$.

\begin{figure}
\includegraphics[width=9.5cm,scale=1.0]{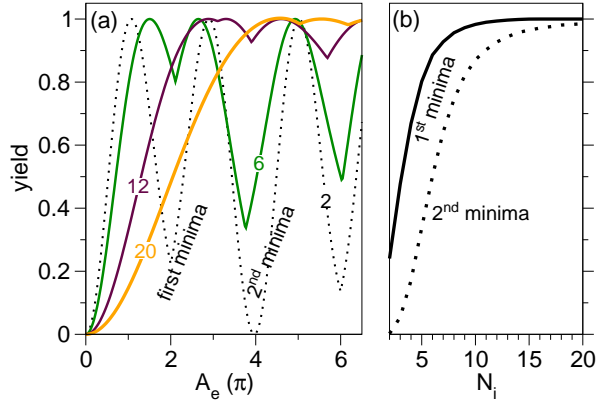}
\caption{Maximum population transfer to the excited manifold at final time
as a function of the extended pulse area for systems with different
number of sublevels $N_i = N_f = N$ and fixed $\Delta E = 0.4 \tau^{-1}$.
In (a) $N = 2$ (dotted line), $N = 5$ (green line), $N = 20$ (orange line).
In (b) we show how the first and second minima of the optimized yield
increases with $N_i$. As shown in (a), the first minima for $N_i = 20$
occurs practically at the first maxima for smaller $N_i$.}
\end{figure}

We now return to the original question: Do quantum structures help to
control the dynamics? 
For population inversion between two manifolds, the problem is easily
controllable by means of proper Rabi oscillations in the simplest case,
with $N_i = N_f = 1$. The existence of a substructure creates an
effective detuning that reduces the oscillations. 
Therefore, the larger the energy spacing $\Delta E$ is, the smaller the
back-effect of the unpopulated states and the more the system resembles
a simple $2$-level system. 
However, by manipulating the state within the initial substructure one can 
regain full Rabi oscillations
and reach a regime where the pulse area
theorem is overcome with almost full
controllability regardless of the pulse area for very large $N_i$.
Fig.3(a) shows how the optimized yield with respect to changes in the
initial state practically achieves full population inversion 
whenever $N_c = N_i$ is large. 
Fig.3(b) shows how the first (and second) minima of the optimal yield
increase with $N_i$.
Only in the degenerate case $N_i$ does not play any role and the dynamics
is less controllable. 
Moreover, since the extended area increases with
the number of sublevels, population inversion can be achieved with 
relatively weak pulses\cite{footnote2}. 
In the given example, with $N_c = N_i = N_f = 20$, full 
inversion is obtained already with a pulse area of ${\cal A} \sim \pi /10$
(see Fig. 4). This is a consequence of parallel transfer.

\begin{figure}
\includegraphics[width=8.0cm,scale=1.0]{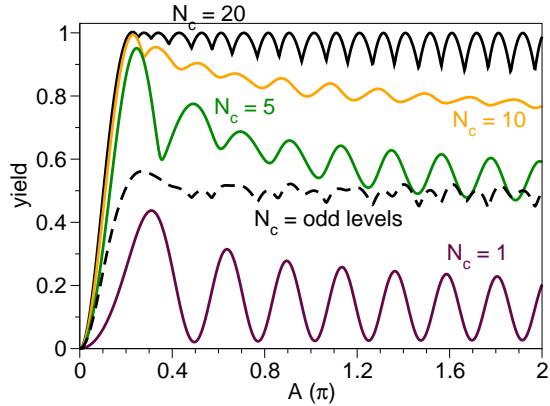}
\caption{Maximum population transfer to the excited manifold at final time
as a function of the pulse area for a system with $N_i = N_f = 20$ sublevels
with control on different subsets of the initial state, $N_c$.}
\end{figure}

On the other hand, one should remember that the ability to optimize 
the yield with increasing $N_i$ (for fixed $\Delta E$) is at the expense
of a finer optimization of the initial state, where the optimal solution
is typically a single one, while the ``robust'' subspace of near-zero 
eigenvalues occupies practically all the space of solutions.
What happens when the dimensionality of the subspace that is 
controlled is smaller than the space that is initially accesible or, 
in other words, how are the solutions deteriorated when $N_c < N_i$? 
In Fig.4 we show the yield of population transfer as a function of the
pulse area for different ``control subspaces''. Now $N_i = 20$ but
the controller has only access to the first $N_c$ sublevels (or to the
odd numbered sublevels, due to {\it e.g.} a unspecified selection rule
or symmetry). The case $N_c = 20$ gives maximum yield while $N_c=1$
implies no control over the initial state. The yields are deteriorated
as the ability to control the system decreases and this effect cannot be
overcome by increasing the pulse area. On the contrary, often best
results are often obtained with ${\cal A}_e = \pi$. 
Moreover, adding external constraints, such as access to only odd
number of levels, returns lower values of the yields.

In summary, we have shown that quantum substructures may hamper the
success of population transfer in multi-level systems. By engineering the 
initial state one can avoid the detrimental effects and partially
correct the Rabi oscillations of the yield forced by the pulse area theorem.
This is not achieved by brute force (increasing the pulse areas) but by
preparing quantum superposition states that 
cancel the detrimental Raman transitions and lead to parallel transfer.
Full population blockade and in fact laser induced transparency 
can also be achieved in similar manners. 
The control over the dynamics increases with the ability to manipulate
every sublevel of the quantum substructure and is substantially reduced
when there is limited control over the sublevels.

\section*{Acknowledgment}
This work was supported by the NRF Grant funded by the Korean government 
(2007-0056343), the International cooperation program (NRF-2013K2A1A2054518), 
the Basic Science Research program (NRF-2013R1A1A2061898),
the EDISON project (2012M3C1A6035358),
and the MICINN project CTQ2012-36184.


\begin{thebibliography}{10}

\bibitem{QC}
S.A. Rice and M. Zhao, Optical Control of Molecular Dynamics 
(John Wiley \& Sons, New York, 2000).
M. Shapiro and P. Brumer, Quantum Control of Molecular Processes 
(Wiley-VCH, 2012).
D. D'Alessandro, Introduction to Quantum Control and Dynamics 
(Chapman \& Hall, 2007).
C. Brif, R. Chakrabarti and H. Rabitz, Adv. Chem. Phys. Vol 148, 1 (2012).

\bibitem{controllability}
G. M. Huang, T. J. Tarn, J. W. Clark, 
J. Math. Phys., 24 , 2608 (1983).
%
V. Ramakrishna, M. V. Salapaka, M. Dahleh, H. Rabitz, A. Peirce, 
Phys. Rev. A, 51, 960 (1995).
%
S. G. Schirmer, H. Fu and A. Solomon, Phys. Rev. A 63, 063410 (2001).

\bibitem{wfcontrol}
G. Turinici, H. Rabitz, Chem. Phys. 267, 1 (2001).
G. Turinici, H. Rabitz, J. Phys. A, 2565 (2003).

\bibitem{molpipulse}
J. S. Melinger, Suketu R. Gandhi, A. Hariharan, J. X. Tull, and W. S. Warren,
Phys. Rev. Lett. 68, 2000 (1992).
J. Cao, C. J. Bardeen, K. R. Wilson, Phys. Rev. Lett. 80, 1406 (1998).
K. Bergmann, H. Theuer, B. W. Shore, Rev. Mod. Phys. 70, 1003 (1998).
N. V. Vitanov, T. Halfmann, B. W. Shore, K. Bergmann, Annu. Rev. Phys. Chem.
52, 763 (2001).
B. M. Garraway, K.-A. Suominen, Contemporary Physics 43, 97 (2002).

\bibitem{Starkshift}
D. Townsend, B.J. Sussman, A. Stolow, 
J. Phys. Chem. A 115, 357, (2011).

\bibitem{Shore}
B. W. Shore, Manipulating Quantum Structures Using Laser Pulses (Cambridge
University Press 2011).

\bibitem{AutlerTownes}
S. H. Autler and C. H. Townes, Phys. Rev. 100, 703 (1955).


\bibitem{IRcontrol}
B. Amstrup and N. E. Henriksen, J. Chem. Phys. 92, 8285 (1992).
N. E. Henriksen, Adv. Chem. Phys. 91, 433 (1995).
S. Meyer and V. Engel, J. Phys. Chem. A 101, 7749 (1997).
N. Elghobashi and L. Gonz\'alez, Phys. Chem. Chem. Phys. 
6, 4071 (2004).

\bibitem{LIT}
O.Kocharovskaya, Ya.I.Khanin, Sov. Phys. JETP 63, 945 (1986).
K.J. Boller, A. Imamoglu, S. E. Harris, Phys. Rev. Lett. 66, 2593 (1991).
Eberly, J. H., M. L. Pons, and H. R. Haq, Phys. Rev. Lett. 72, 56 (1994).

\bibitem{footnote2}
Obviously the minimization of the area of the pulse $\Omega$ is at
the expense of previous pulses that are needed to prepare the initial state.

\end{thebibliography}
\end{document}